\documentclass[conference]{IEEEtran}

\ifCLASSINFOpdf
   \else
    \usepackage[dvips]{graphicx}
\fi
\begin{document}
%
\title{Terahertz Parametric Gain in Semiconductor Superlattices}


\author{\IEEEauthorblockN{Timo Hyart}
\IEEEauthorblockA{Department of Physical Sciences,\\ University of
Oulu, Finland\\ Email: timo.hyart@oulu.fi} \and
\IEEEauthorblockN{Alexey V. Shorokhov} \IEEEauthorblockA{Institute of Physics and Chemistry,\\
Mordovian State University,\\ 430000 Saransk, Russia} \and
\IEEEauthorblockN{Kirill N.
Alekseev} \IEEEauthorblockA{Department of Physical Sciences,\\
University of Oulu, Finland\\ Email: kirill.alekseev@oulu.fi} }


%


\maketitle

\begin{abstract}
We  consider  a high-frequency response of electrons in a single
miniband of superlattice subject to dc and ac electric fields.
Action of ac electric field causes oscillations of electron's
effective mass in miniband, which result in a parametric
resonance. We have established a theoretical feasibility of
phase-sensitive parametric amplification at the resonance. The
parametric amplification does not require operation in conditions
of negative differential conductance. Therefore a formation of
destructive domains of high electric field inside the superlattice
can be prevented. Here we concentrate on
the parametric up- and down-conversion of electromagnetic
radiation from available frequencies to desirable THz frequency
range.
\end{abstract}


%

\IEEEpeerreviewmaketitle

\section{Introduction}
Terahertz spectral range of electromagnetic radiation is located
between optical and microwave frequencies. It is the borderline of
optics and electronics, and there still exists a
technological gap in construction of compact and reliable sources,
amplifiers and detectors. Semiconductor superlattices
have been proposed as a candidate for active material in such kind
of devices \cite{esaki70}.
%
Realization of a regenerative superlattice THz
amplifier/oscillator is complicated by an existence of electric
instability, which leads to a formation of destructive high-field
domains inside the nanostructure.

Recently we theoretically established the feasibility of
regenerative parametric amplification in superlattices due to the
parametric resonance caused by the oscillations of electron's
effective mass in miniband \cite{Hyart2007}.
We showed that the parametric amplification does not require
operation in conditions of negative differential conductance
\cite{Hyart2006} and thus, at least for a moderate doping, the
formation of destructive high-field domains inside the
superlattice can be prevented \cite{Alekseev2006}.
The superlattice parametric amplifier does require a relatively
strong ac pump. Currently, suitable powerful sources of coherent
radiation exist at the low frequencies of about 100 GHz and at some
particular high frequencies of several THz \cite{Tonouchi}.

After a short description of the amplification schemes, we study the
parametric amplification and generation at even harmonics and
high-order half-harmonics of the pump frequency, as well as
amplification and generation with frequency halving. We focus on the
parametric amplification and generation of desirable THz radiation
(200 GHz - 1 THz) with the use of available sources of low- and
high-frequency radiation.

\section{Amplification schemes}

We employ the semiclassical approach based on a solution of
Boltzmann transport equation \cite{Hyart2007}. We consider
electron dynamics in a single superlattice miniband subject to the
pump field consisting of dc bias and strong ac field
\begin{equation}
E_{\rm p}=E_{\rm dc}+E_0\cos\omega t
\end{equation}
and a weak, phase-shifted probe field
\begin{equation}
E_{\rm pr}=E_1 \cos(\omega_1 t+\phi).
\end{equation}
There naturally arises two distinct schemes of parametric
amplification of the probe field without corruption from generated
harmonics: Parametric amplification at even harmonics
$\omega_1=:2\omega, 4\omega,...$ in the unbiased superlattice
($E_{\rm dc}=0$) and amplification at half-harmonics
$\omega_1=:\omega/2, 3\omega/2,...$ in the biased superlattice.

For both these schemes absorption of the probe field can be
represented as a sum of phase-dependent coherent and
phase-independent incoherent components. The incoherent component of
absorption can be interpreted as a free-carrier absorption modified
by the pump field, while the coherent contribution is caused by the
periodic variation of effective mass and a specific quantum
inductance. The parametric gain has a maximum for some optimal value
of the relative phase $\phi=\phi_{\rm opt}$ \cite{Hyart2007}.

In our calculations of magnitude of parametric gain we used the
following parameters of typical GaAs/AlAs superlattice: superlattice
period $d=6 \ \rm{nm}$, miniband width $\Delta= 60 \ \rm{meV}$,
scattering time $\tau=200 \ \rm{fs}$, density of electrons
$N=10^{16} \ \rm{cm}^{-3}$ and temperature $300 \ \rm{K}$. For such
superlattice the Esaki-Tsu critical field $E_c$  \cite{esaki70}
is approximately $5.5$ kV/cm.

\section{Parametric up-conversion}

Both schemes, parametric amplification at even harmonics and
half-harmonics, can be used for up-conversion of electromagnetic
radiation from microwave frequencies to the THz frequency range
\cite{Shorokhov06}.
%

\begin{figure}
  \includegraphics[scale=0.6]{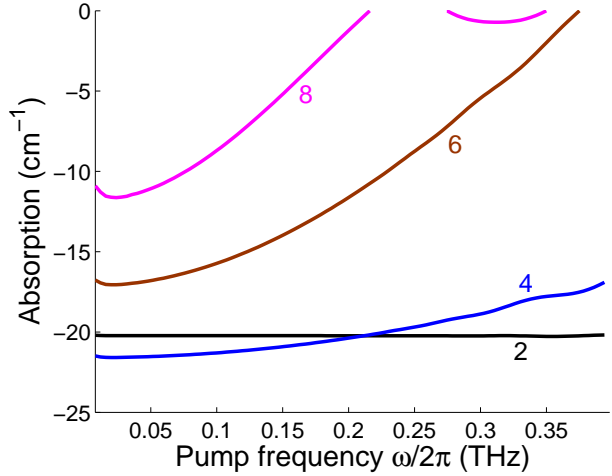}
  \caption{Magnitude of the small-signal gain at even harmonics $\omega_1 =n \omega$ ($n=2,4,6,8$)
  as a function of pump frequency $\omega$ for the fixed amplitude $E_0=6.5 E_c$ and $\phi=\phi_{\rm
  opt}$.}
  \label{fig1}
\end{figure}

As can be seen from Fig. \ref{fig1}, the magnitude of the gain is
still significant even for $\omega_1=6\omega$ with $\omega/2\pi$
being around 100 GHz. In particular, this enables amplification of
probe field with the frequency of about $1$ THz by using a pump
field of frequency 170 GHz. By using the pump field of a higher
frequency, amplification of $1$ THz probe field can be also achieved
at lower frequency ratios, like  $\omega_1/\omega=2$ or
$\omega_1/\omega=4$.

\begin{figure}
  \includegraphics[scale=0.6]{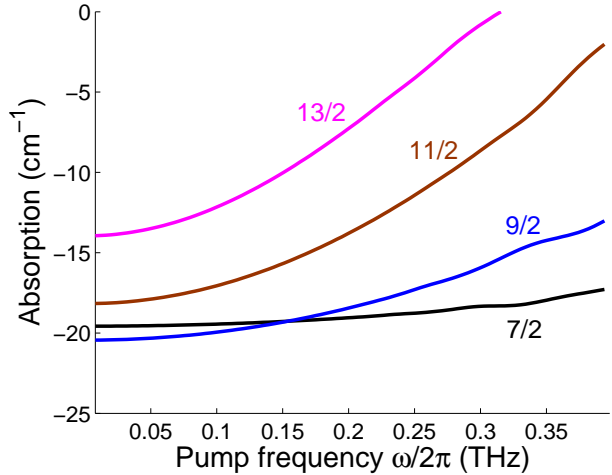}
  \caption{Magnitude of the small-signal gain at half-harmonics $\omega_1
  =: 7\omega/2, 9\omega/2, 11\omega/2, 13\omega/2$
  as a function of pump frequency $\omega$ for the pump amplitude $E_0=6.5 E_c$, dc bias $E_{\rm dc}=E_c$
  and $\phi=\phi_{\rm opt}$.}
  \label{fig2}
\end{figure}

As can be seen from Fig. \ref{fig2}, the magnitude of the gain at
high-order half-harmonics in biased superlattices is no less than
the gain at high-order even harmonics in unbiased superlattices.
Thus, amplification of probe field of 1 THz can be achieved also in
the case, where the probe frequency is a half-harmonic of the pump
frequency.

\section{Parametric down-conversion}

\begin{figure}
  \includegraphics[scale=0.6]{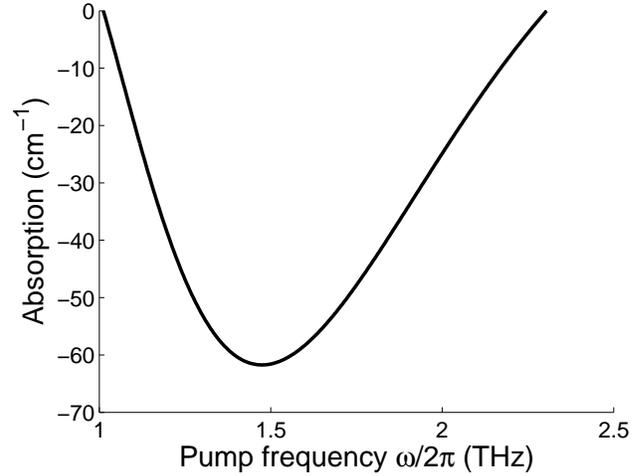}
  \caption{Magnitude of the small-signal gain at $\omega_1 =\omega/2$
  as a function of pump frequency $\omega$ for the pump amplitude $E_0=2.6
  E_c$,  bias $E_{\rm dc}=0.8 E_c$
   and $\phi=\phi_{\rm opt}$.}
  \label{fig3}
\end{figure}

The parametric amplification at half-harmonics, can be also used for
the parametric down-conversion. Fig. \ref{fig3} shows that large
gain at $\omega_1 =\omega/2$ can be achieved with the pump field of
modest amplitudes. Therefore, employing a suitable pump field with
frequency 1-2 THz -- that can be generated with the help of modern
quantum cascade lasers or devices based on difference-frequency
optical mixing \cite{Tonouchi} -- a probe field with frequency 0.5-1
THz can be amplified in superlattice.

\section{Conclusion}

Our theory showed that the parametric resonance in superlattice
miniband can be used for the phase-sensitive parametric
amplification at THz frequencies. Moreover, all the requirements for
frequencies and strengths of the pump fields can be already
fulfilled at the present state of microwave and terahertz
technologies. Therefore, superlattice parametric devices can
potentially form a basis for future amplifiers operating in THz
frequency domain.

In conclusion, we would like to note that it is also possible to
achieve phase-insensitive amplification in conditions of suppressed
electric domains for incommensurate pump and probe frequencies
\cite{Hyart2007b}. In this case, high-frequency gain in superlattice
resembles Bloch gain, i.e. it occurs due to scattering-assisted
quantum transitions.


\section*{Acknowledgment}

This work was partially supported by Academy of Finland (109758
and 101165), Emil Aaltonen Foundation, grant of President of
Russia for Young Scientists (MK-4804.2006.2) and AQDJJ Programme
of ESF.




\begin{thebibliography}{1}

\bibitem{esaki70}
L. Esaki and R. Tsu, IBM J.~Res.~Dev. \textbf{14}, 61 (1970).

\bibitem{Hyart2007} T. Hyart, A. V. Shorokhov and K. N. Alekseev,
Phys. Rev. Lett. \textbf{98}, 220404 (2007).

\bibitem{Hyart2006} T. Hyart, N. V. Alexeeva, A. Lepp\"{a}nen and K. N. Alekseev, Appl.
Phys. Lett. \textbf{89}, 132105 (2006).

\bibitem{Alekseev2006} K. N. Alekseev, M. V. Gorkunov, N. V. Demarina, T. Hyart, N. V.
Alexeeva and A. V. Shorokhov, Europhys. Lett. \textbf{73}, 934
(2006).

\bibitem{Tonouchi} M. Tonouchi, Nature Photonics, \textbf{1}, 97
(2007).

\bibitem{Shorokhov06}
A. V. Shorokhov and K. N. Alekseev, Physica E \textbf{33}, 284
(2006).

\bibitem{Hyart2007b} T. Hyart, K. N. Alekseev, A. Lepp\"{a}nen, E.
V. Thuneberg, in preparation.

\end{thebibliography}
%

\end{document}